\documentclass[twocolumn]{article}
\usepackage[utf8]{inputenc}
\usepackage{graphicx}
\usepackage{hyperref}
\usepackage{geometry}
\usepackage[bottom]{footmisc}
\usepackage{lettrine}
\usepackage{tabularx}
\usepackage{caption,newfloat}
\DeclareFloatingEnvironment[
  fileext=lob,
  listname={List of Boxes},
  name=Box,
]{BOX}
\usepackage{afterpage}
\usepackage{float}
\usepackage{hyperref}
\usepackage{amssymb}
\usepackage{mathtools}
\usepackage{textgreek}
\usepackage{listings}
\usepackage[explicit]{titlesec}
\usepackage[T1]{fontenc}
\usepackage{beramono}
\usepackage{listings}
\usepackage[usenames,dvipsnames]{xcolor}
\lstdefinelanguage{Julia}%
  {morekeywords={abstract,break,case,catch,const,continue,do,else,elseif,%
      end,export,false,for,function,immutable,import,importall,if,in,%
      macro,module,otherwise,quote,return,switch,true,try,type,typealias,%
      using,while},%
   sensitive=true,%
   alsoother={\$},%
   morecomment=[l]\#,%
   morecomment=[n]{\#=}{=\#},%
   morestring=[s]{"}{"},%
   morestring=[m]{'}{'},%
}[keywords,comments,strings]%
\begin{document}
%for space after section title 
 \titlespacing\section{0pt}{12pt plus 4pt minus 2pt}{0pt plus 2pt minus 2pt}

\twocolumn[
\begin{center}
{ \huge Julia for Biologists}
\end{center}
\begin{flushleft}

Elisabeth Roesch\footnotemark, 
Joe G. Greener\footnotemark, 
Adam L. MacLean\footnotemark, 
Huda Nassar\footnotemark, 
Christopher Rackauckas\footnotemark, 
Timothy E. Holy\footnotemark, 
Michael P.H. Stumpf\footnotemark{*}
\end{flushleft}
\begin{flushleft}
\bf Increasing emphasis on data and quantitative methods in the biomedical sciences is making biological research more computational. Collecting, curating, processing, and analysing large genomic and imaging data sets poses major computational challenges, as does simulating larger and more realistic models in systems biology. Here we discuss how a relative newcomer among computer programming languages -- Julia -- is poised to meet the current and emerging demands in the computational biosciences, and beyond. Speed, flexibility, a thriving package ecosystem, and readability  are major factors that make high-performance computing and data analysis available to an unprecedented degree to ``gifted amateurs''. We highlight how Julia's design is already enabling new ways of analysing biological data and systems, and we provide a, necessarily incomplete, list of resources that can facilitate the transition into the Julian way of computing.

\end{flushleft}
]
\setcounter{footnote}{1} % Lisi
\footnotetext{School of Mathematics and Statistics, University of Melbourne, 813 Swanston Street, Parkville VIC 3010, Australia. Melbourne Integrative Genomics, University of Melbourne, 30 Royal Parade, Parkville VIC 3052, Australia.}

\setcounter{footnote}{2} % Joe
\footnotetext{Department of Computer Science, University College London, Gower Street, London, WC1E 6BT, UK.}

\setcounter{footnote}{3} % Adam
\footnotetext{Department of Quantitative and Computational Biology, University of Southern California, 1050 Childs Way, Los Angeles, CA 90089, USA.}

\setcounter{footnote}{4} % Huda
\footnotetext{RelationalAI, Inc. 2120 University Ave, Berkeley, CA, 94794, USA.}

\setcounter{footnote}{5} % Chris
\footnotetext{Department of Mathematics, Massachusetts Institute of Technology, 182 Memorial Dr, Cambridge, MA 02142, USA. Julia Computing, 240 Elm Street, 2nd Floor, Somerville, Massachusetts 02144, USA. Pumas-AI, 14711 Kamputa Drive, Centerville, VA 20120, USA.}

\setcounter{footnote}{6} % Tim
\footnotetext{Departments of Neuroscience and Biomedical Engineering, Washington University in St.~Louis, 660 S.~Euclid Ave., St.~Louis, MO 63110, USA.}

\setcounter{footnote}{7} % Michael
\footnotetext{*Corresponding author: mstumpf@unimelb.edu.au. School of Mathematics and Statistics, University of Melbourne, 813 Swanston Street, Parkville VIC 3010, Australia. School of BioSciences, Biosciences 4, The University of Melbourne, Royal Parade, Parkville VIC 3052, Australia. Melbourne Integrative Genomics, University of Melbourne, 30 Royal Parade, Parkville VIC 3052, Australia.}

%\textbf{\textit{Computers are tools.}} 
\lettrine[findent=2pt]{\color{ForestGreen}{\textbf{C}}}{ }
omputers are tools. Like pipettes or centrifuges, they allow us to perform tasks more quickly or efficiently; and like microscopes, NMR or mass-spectrometers, they allow us to gain new, more detailed insights into biological systems and data. Computers also allow us to define, simulate and test mathematical models of biology. As computational power evolved, solving biological problems computationally became possible, then popular, and eventually, necessary \cite{Tomlin2007}. Entire fields such as computational biology and bioinformatics emerged. Without computers, the reconstruction of structures from X-ray crystallography, NMR, or cryo-EM methods would be impossible. The same goes for the genome project~\cite{genome-project}, which used computer programs to assemble and analyze the DNA sequences generated; and, to this day, computer programs continue to enable new science by analyzing data from the genome project. More recently, vaccine development has benefited immensely from recent advances in algorithms, software, and computer hardware~\cite{ROBSON2020103670}.
\par 
%\textbf{\textit{Programming languages are also tools.}} 
Programming languages are also tools. They provide the bridge between hypothesis or model formulation and computational power. Programming languages make it possible to instruct computers to  run algorithms, for example for the analysis of biological data. Some languages are very good at specific tasks -- think Perl for string processing tasks; or R for statistics and data analysis -- whereas  others -- including C, C++, and Python -- have been used with success across many different domains. In biomedical research the prevailing languages have arguably been R \cite{Seefeld2007} and Python \cite{Ekmekci2016}. Much of the high-performance backbone supporting computationally intensive research, hidden from most users, however, continues to rely on C/C++ or Fortran. Many computationally intensive studies are designed in a way where an initial first draft is coded in R, Python or Matlab (first language), and subsequently translated into C/C++ or Fortran (second language) for performance reasons. This is known as the two-language problem.
\par 
%\textbf{\textit{The two language problem.}} 
While this two-language approach has effectively facilitated and sped up scientific discovery in many instances, one can imagine instances where this model has been limiting. When moving a certain implementation from one programming language to a second, faster, programming language, straightforward ``verbatim" translation may not be the optimal route: the faster language  (such as C/C++ or Fortran) often provides the programmer with a much higher margin of autonomy, such as the ability to choose how memory is accessed or allocated or to employ slightly more sophisticated data structures \cite{Perkel2020}. Exploiting these gains may require a complete rewrite of the algorithm \cite{Ripley:1987ud,nazarathy2021statisticsjulia} to ensure  faster implementations, faster scaling, or potentially better packaged code. This requires expertise across both languages, but also rigorous testing of the code in both languages.  
%
%one can miss the opportunity to model the problem slightly differently which may allow even faster implementations, faster scaling, or potentially better packaged code that can be used to serve many future scientific discoveries. This does not happen only because the second language is often faster -- it is also because the second language 
\par 
\begin{figure*}[pt]
    \centering
    \includegraphics[width=\textwidth]{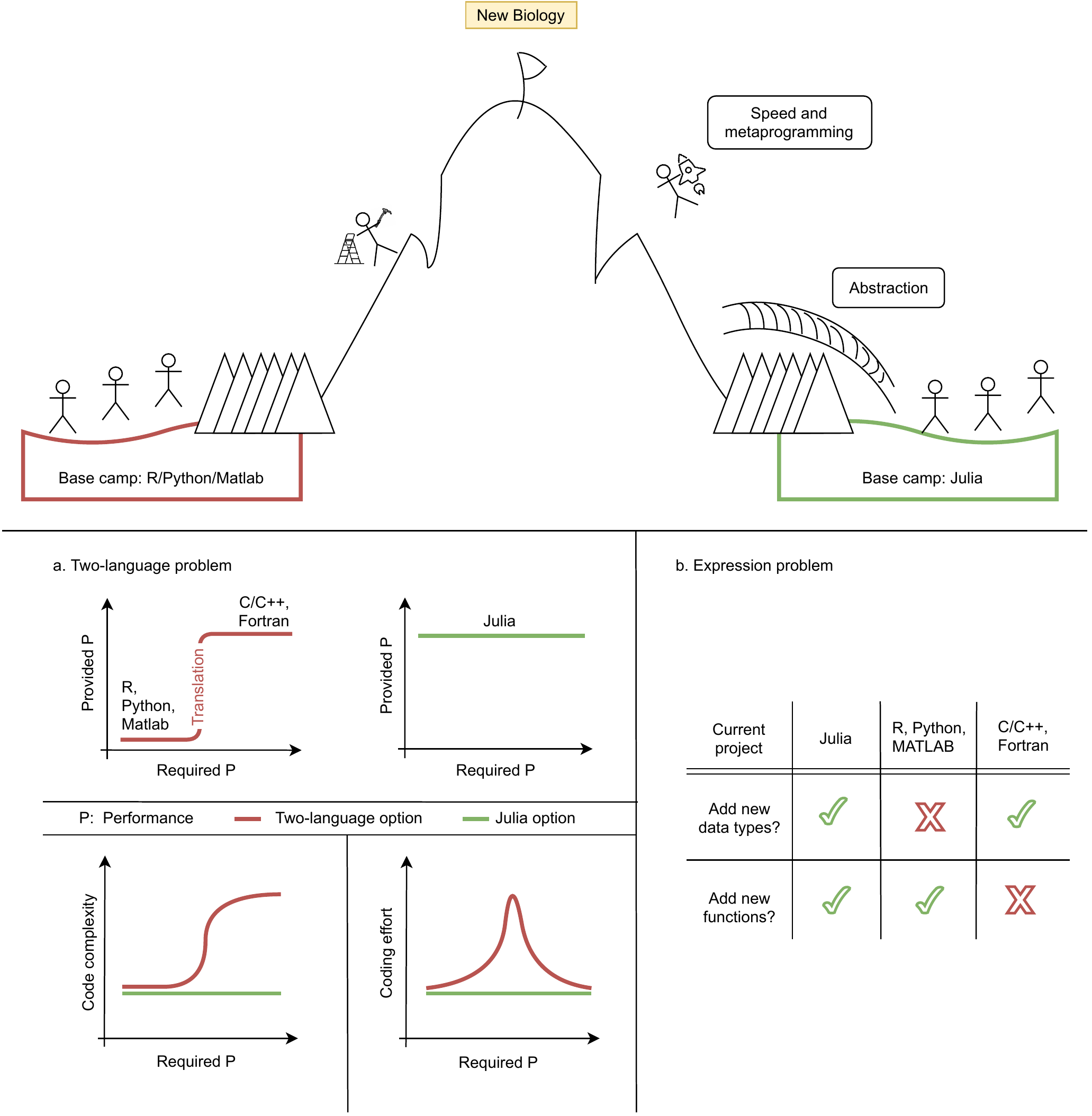}
    \caption{Julia is a tool for the biologist to discover new science. In the biological sciences, the most obvious alternative to the programming language Julia is R, Python or Matlab. Here, we contrast the two potential pathways to new biology with a mountaineering excursion: The top of the mountain represents "New Biology". There are two potential base camps for the ascent: Base camp 1 (left, red) is "R/Python/Matlab". Base camp 2 (right, green) is "Julia". To get to the top, the mountaineer -- representing the biologist -- needs to overcome certain obstacles such as a glacier and a chasm. They represent research hurdles such as large and diverse data sets or complex models. Starting at the "Julia" base camp, the mountaineer has access to efficient and effective tools such as a bridge over the glacier and a rocket to simple fly over the chasm. They represent Julia's top three language design features: Abstraction, speed and metaprogramming. With these tools, the journey to the top of the mountain becomes much easier for the excursionist. In other words, Julia allows the biologist to not be hold back by problems such as the "Two-language problem" and the "Expression problem" and therefore Julia helps the biologist to discover new science.}
    \label{fig:cartoon}
\end{figure*}

\begin{table*}[t]

\begin{tabularx}{\textwidth}{|m{2.1cm}|m{14.7cm}|}
\hline
    1. Language design & \begin{itemize}
            \setlength\itemsep{-0.1em}
            \item Julia is user friendly. $\rightarrow$ It is easy to code.
            \item Julia is a high performance language. $\rightarrow$ It is fast.
            \item Julia offers a high level of abstraction.  $\rightarrow$ It is flexible.
            \item Julia can be used for metaprogramming. $\rightarrow$ It can code {\it automatically}.
            \item Julia is not only good in one area but in many.  $\rightarrow$  It enables "one language" projects.
        \end{itemize}\\
        \hline
    2. Low barrier to entry &
    
        \begin{itemize}
        \setlength\itemsep{-0.1em}
            \item Easy to learn due to intuitive semantic and easy to read syntax.
            \item Accessible via various interfaces, REPL, IDE, or Jupyter notebook.
            \item Existing non-Julia code can be easily integrated into new Julia projects via language specific packages (Figure \ref{fig:packages}, Integration of non-Julia code). 
        \end{itemize}
        \\
        \hline
    3. Additional reasons
        &
        \begin{itemize}
        \setlength\itemsep{-.1em}
            \item Julia is free, open-source and hosted on GitHub.
            \item Julia offers (generally) excellent documentation, tutorials, and help available directly from active and welcoming community members via various communication channels such as Slack, Discourse, Twitter or Zulip.
            \item Julia's package ecosystem provides functionality for a wide range of oft-performed tasks in computational biology research (Figure \ref{fig:packages}, Table \ref{tab:packages}).
            \item Julia code is smoothly extendable which enables and encourages easy contributions and collaborations to/with existing projects,  as well as
            writing, integrating and sharing new, user specific packages.
        \end{itemize}\\
\hline
\end{tabularx}
\caption{As Julia is a relatively young language, it is save to assume that the majority of biologists are not using Julia yet. In this table we present arguments that make Julia a good programming language for biologists.}
  \label{tab:switch}
\end{table*}
%\textbf{\textit{Computational biology and the two language problem.}} 
As the field of computational biology evolves, the two-language approach is still surprisingly persistent: higher-level languages (R, Python) are used for algorithm development and preliminary analyses. The better performing languages (C/C++, Fortran, or CUDA) are used in specific instances and with considerable additional effort if the computational burden becomes overwhelming. This is questionable from an efficiency perspective but also problematic in regards to the correctness and accuracy as the translation step leaves room for mistakes and misinterpretation; especially if the translation is done by someone who is not a domain expert \cite{Carey2018}. This issue will only become more problematic as the demand for high performance code with good scalability continues to  increase to meet the demands of bigger data sets and more detailed models.  However, this also makes it very evident, that the accessibility and ease-of-use aspect of the programming language is of considerable importance to biomedical scientists, and, implicitly, the progress of biomedical research.
\par 
%\textbf{\textit{Julia challenges the two language problem.}} 
Julia \cite{Bezanson2017} is a relatively new programming language, and the main advantage for biologists is that it challenges the two-language concept, by being a language that ``looks like Python and runs like C". Users do not have to choose between ease-of-use and high performance -- Julia is built to be easy to program in and fast \cite{Perkel2019}. This and the growing ecosystem of state-of-the-art application packages and introductions \cite{nazarathy2021statisticsjulia,Lauwens:2021uj} make it an attractive choice for biologists.
\par
Biological systems and data are multifaceted by nature, and to describe them, or model them mathematically, requires flexibility of a programming language to handle of connect different types of highly structured data, see Figure~\ref{fig:cartoon}. There are three hallmarks of the language which make it particularly suitable to meet the current and emerging demands of biomedical scientists that we discuss here in some detail: speed, abstraction and metaprogramming. We  discuss each language feature and its biological relevance in the context of concrete examples. And we provide a basic toolset and tips to get started with Julia, see Table~\ref{tab:switch}. Further  supporting online material as well as code examples are provided in \href{github.com/ElisabethRoesch/Perspective_Julia_for_Biologists}{\textit{github.com/ElisabethRoesch/Perspective\_Julia\_for\_ Biologists}}.
 
 %
% there is just no question on the immense potential of an elegant mathematical modeling framework in the hands of a biological domain expert with an accessible and efficient connection to computational power in order to bring their theoretical models to life. With a programming language like Julia, code does not only run a bit smoother and quicker but it actually enables biologists to substantially discover wider scopes of their models and research. 
\par 
%\textbf{\textit{Many biological problems could benefit from a unified language.}} 
%For example, in a data-driven model selection and parameter estimation problem in an analysis framework of pathway dynamics based on transcriptomic single cell data, the computational power needed is enormous. However, the underlying system and chosen models are complex and rely heavily on biological domain expertise. Understanding the biological assumptions in order for specific simulation algorithms to be applicable is difficult to communicate. The old-fashioned approach of an initial biological draft in R, Python or Matlab and a subsequent delegation to a software developer in C/C++ or Fortran does clearly not provide enough flexibility to accommodate for this type of project. Another example are big computational biology models such as whole cell models. Again, the computational power needed is enormous and the modeling framework extremely complex and relies heavily on biological domain expertise. On top of this, models of this scope do rely on automated coding, i.e. metaprogramming. There are not many languages that could potentially accommodate this level of abstraction; if we consider accessibility a non-negotiable, there is no comparable language to Julia.
\par
 %\textbf{\textit{Julia is an easy language to switch to.}} 
 %Usually, switching to a new machine or algorithm often entails learning how to run the new machine or implementing a new algorithm. This is not exactly the same for programming languages, as the cost is often not just learning the new programming language, but also the often current reliance on existing packages of other programming languages or an existing large codebase. Nevertheless, in Julia's case, the benefits arguably outweigh the cost of this switch. This is simply because if you program in Julia, you can continue to call your favorite functions from a certain package without too much extra hassle. This makes it easy to enjoy all of Julia's benefits immediately without a large backlog of trying to translate existing code. Julia can provide a paradigm shift in how we think about computational problems whether by allowing us to increase the scale or the complexity of problems we are trying to solve, and in turn, push the boundaries of science.

\section*{Speed}
\label{sec:speed}
The speed of a programming language is not just a matter of convenience that allows us, for example, to finish analyses more quickly; it also enables new science and better science. Speed is important when analysing large data sets \cite{Marx2013, Chan2017} that are becoming the norm in many areas of biology \cite{Svensson2018}. Slow operations might not hinder scientific discovery when performed a small number of times. However, when performed repeatedly on large data set, the speed of a programming language can become the limiting factor for new discoveries (See: Speed Example 1). Similarly, simulating large and complex computational models is only possible with fast implementations (See Speed Example 2); and digital twins \cite{Bjornsson:2019vu,Laubenbacher:2021vy} in precision medicine are useless without fast and convenient computation. 
\par 
The speed of the programming language also determines how extensively we can test statistical analysis or simulation algorithms before using them on real data. Thorough testing of a new statistical algorithm can be expected to be around 2-3 orders of magnitude more costly in computational terms than a single ``production run" \cite{Chan2017}. Furthermore, the quality of approximations also depends on many factors (e.g. number of tested candidates \cite{Tankhilevich2020, Innes2018} and grid step sizes \cite{Rackauckas2017}) and faster code enables better analysis. 
\par 
In the following we use two examples of Julia in Biology to highlight some of Julia's speed features and showcase how they enable new biology. We also provide some technical insights into the design features underlying Julia's speed \cite{Sengupta:2019uj}.   
\begin{figure*}[pt]
    \centering
    \includegraphics[width=\textwidth]{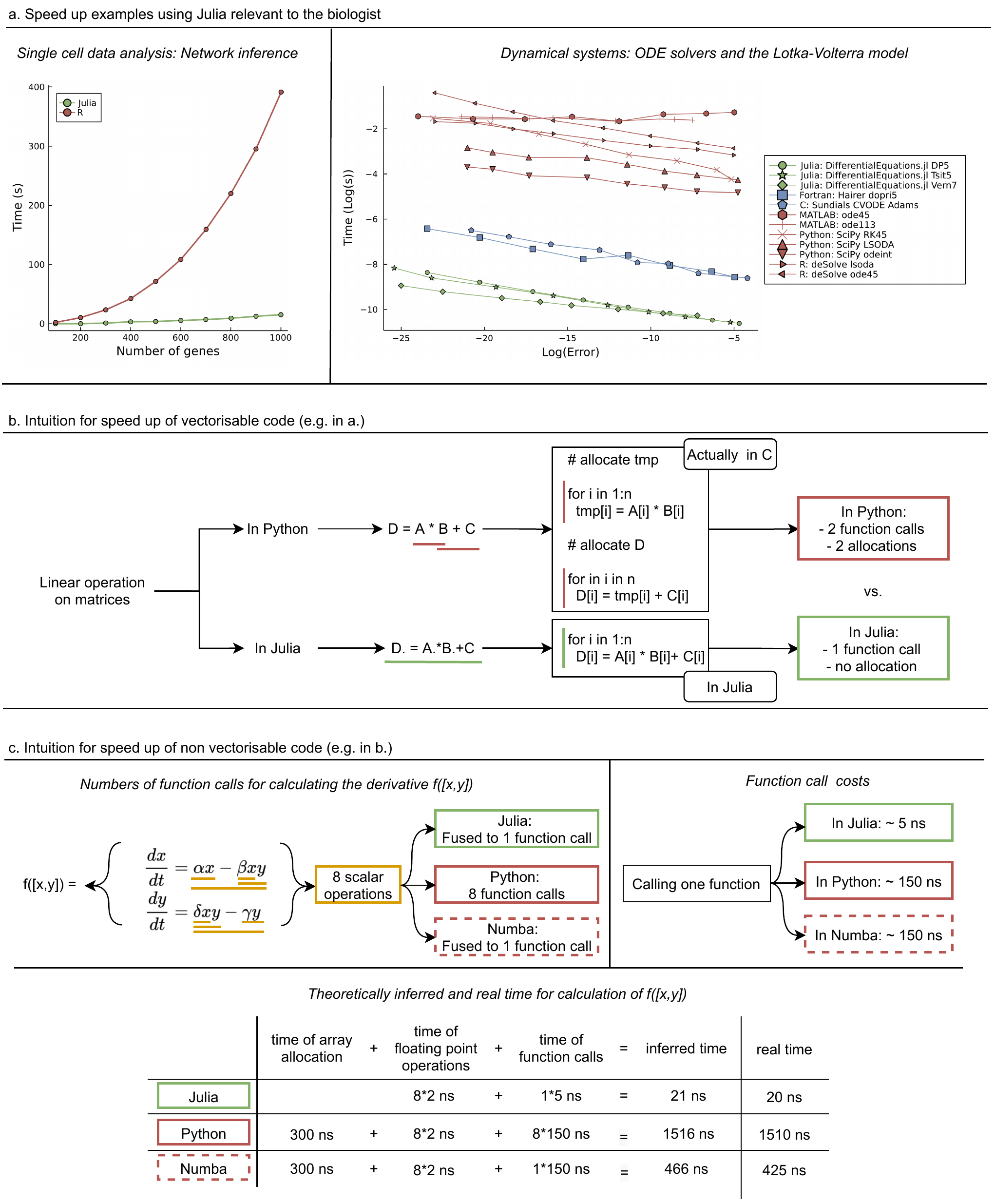}
    \caption{Julia's speed feature. (a) Examples relevant to the biologist. Left: Time to calculate mutual information for all possible pairs of genes of a single cell data \cite{Chan2017}. Right: Benchmark of ODE solvers (More systems in \cite{SciMLBenchmark}). (b) Illustration of speed-up of vectorisable code (as in (a)). (c) Intuition for speed up of non vectorisable code (as in b).}
    \label{fig:speed}
\end{figure*}
\paragraph{Example 1: Fast Network Inference from Single Cell Data.}
\label{subsec:fast_network_inference}
In single cell biology, we can now measure the expression levels of tens of thousands of genes in tens of thousands of cells. Increasingly we are even able to do this with high spatial (that is sub-cellular) resolution \cite{Chen2017}. But searching for patterns in complex and large data-sets can be computationally expensive: even apparently simple tasks, such as calculating the mutual information (MI) \cite{McMahon:2014cs} across all pairs of genes in a large data set can quickly become impossible.  
\par 
The inference of gene regulatory networks (GRN) from single cell data is a statistically demanding task, and one where Julia's speed can help. This is illustrated by the analysis of  Chan et {\it al.} \cite{Chan2017} where information theoretical measures were used to infer GRNs from transcriptomic single cell data of qPCR experiments on megakaryocyte-erythroid progenitor cells during human hematopoiesis \cite{Psaila2016}, early embryonic development \cite{Guo2010}, and embryonic hematopoietic development \cite{Moignard2015}. The MI has to be calculated for gene pairs; but a multivariate information measure, {\em partial information decompositio} (PIDC) is also considered to separate out direct and indirect interactions \cite{McMahon:2014cs}, and this requires looking at all gene triplets \cite{Stumpf:2017fa}. 
\par
The run-time of algorithms implemented in the Julia package {\tt InformationMeasures.jl}  can be compared to the popular R package {\tt minet} \cite{Meyer2008} for different number of genes and numbers of cells (Figure \ref{fig:speed}(a) left). For small numbers of genes, differences are significant but not prohibitive: inferring a network with 100 genes takes around 0.3 seconds in Julia compared to 1.5 seconds in R; but already for 1,000 genes the inference times differ substantially: 17 seconds in Julia and 390 seconds ($>$20-fold difference) in R; for -- by today's standards small -- datasets with 3,500 genes and 600 cells R needs over 2.5 hours, compared to Julia's 134 seconds ($\approx$ 64-fold difference); and in real-world applications \cite{Stumpf:2017fa} 400-fold speed differences are possible. Here we are reaching the threshold of what can be tested and evaluated rigorously in many highlevel languages. Overall, multivariate information measures would almost certainly be unfeasible in pure R or Python implementations. 
\par
The reason for this performance difference is Julia's ability to optimize ``vectorizable'' code; cf. Figure \ref{fig:speed}(b) \cite{Sengupta:2019uj}. Users of languages like Python and  R are familiar with vectorized functions, such as maps and element-wise operations. Julia gains further performance improvements by combining its Just-in-Time (JIT) compilation with vectorized functions via a trick known as operator fusion. When writing a chain of vector expressions, like $D=A*B+C$ (where $A, B, C$ and $D$ and $n$-dimensional vectors), libraries like NumPy call optimized code that works under the hood, and which is generally written in another, faster language like C/C++, and these operations are computed sequentially. For this example, $A*B$,  C code is called to produce a temporary array, $tmp$, and then $tmp+C$ is again evaluated using C code to produce the desired $D$. Allocating the memory for the temporary intermediate, $tmp$, and the final result $D$  is $O(n)$ (which means that the time it takes to complete the computation increases with approximately with $n$, the length of the vectors), and scales proportionally to the compute cost; thus no matter what the size of the vectorization is, there is a major unavoidable overhead. Julia uses the ``$.$'' operator to signify element-wise action of a function, and therefore the equation can be written as $D.=A.*B.+C$. When the Julia compiler sees this, so-called {\em broadcast}, expression it fuses all nearby dot-operations into a single function, and JIT compiles this function into a loop. In concrete terms, NumPy makes two function calls and spends time generating two arrays, whereas Julia makes a single function call and reuses existing memory. This and similar performance features are now  leading package authors of statistical and data science libraries to recommend calling into Julia for such operations, such as the recommendation by the principal author of the R {\tt lme4} linear mixed effects library to use {\tt JuliaCall} to access {\tt MixedModels.jl} in Julia (both written by the same author) for an approximately 200x acceleration\cite{Bates:2018aa}. 

\paragraph{Example 2: Accelerating Dynamical Systems Modeling in Systems Biology and Pharmacology.}
Systems biology and related fields, including quantitative systems pharmacology (QSP), are also benefitting from Julia's speed.  Modeling and simulation are transforming the drug discovery pipeline, lowering the risk of failed trials, and allowing efficiency gains in drug development and substantial financial savings in the drug development process \cite{efpia2016good}. However, even with these successes most trials do not undergo in-depth preclinical analysis. The major reason why is time: any delay in the start of the clinical trial increases the overall cost. Improvements in QSP can remedy this situation. 
\par 
Solving large systems of ordinary differential equations (ODEs) (and increasingly also stochastic dynamical systems)  lies at the core of these modelling studies. We typically have nonlinear functions, $f$, and solving them in high-level languages such as R, Python or Matlab can be slow. Therefore solver libraries are often written in a faster language, such as C/C++ or Fortran. The limiting factor then is the user's non-linear set of equations, $f$. In languages like Python or R, there is a high function call overhead: every operation that is called is more expensive than in a fast language (approximately 150-350ns per call \cite{Nunez-Iglesias:2018aa} while the function calls can take approximately 5ns in Julia or C). Scalar operations, like evaluating a Hill kinetic function $[A]^\prime = \frac{[B]^n}{\omega + [B]^n}$, can  take microseconds instead of nanoseconds, see Figure \ref{fig:speed}. ``Vectorization'', as recommended in languages such as Python or R, packs more floating point operations into each C function call and can help to speed this up somewhat. Even accelerators like Numba still require a context change from Python to the compiled C function, which can hamper performance, especially for sparse reaction networks. Furthermore, vectorization requires a certain level of regularity and simplicity in the equations, and the nonlinear systems typically found in biology can be anything but simple; therefore traditional interpreted languages will always tend to perform poorly for nonlinear models.
\par 
When solving an ODE, the function $f$ is called thousands or millions of times, exacerbating this difference. Figure \ref{fig:speed} showcases some examples of biological models where such simulations are 50x-400x faster than those using leading packages in R and Python. In a typical preclinical drug development pipeline this has led to 175-fold acceleration of QSP model analysis once the model had been translated from a combination of MATLAB and C code into Julia \cite{rackauckas_ma_rieger_sher_allen_ivaturi_shah_musante_2020}. Julia's speed enables more efficient clinical trial analyses and its libraries have been shown to be even faster than commonly used Fortran libraries in this domain of ODE modeling \cite{rackauckas2020accelerated}.

%Using these tools, a team of MIT researchers with Pfizer's quantitative systems pharmacology team recently demonstrated a 175x acceleration to a production QSP model analyses in the preclinical drug development pipeline by translating the model from a combination of C and MATLAB to pure Julia \cite{rackauckas_ma_rieger_sher_allen_ivaturi_shah_musante_2020}. 
% Modern tools for clinical trial analyses, such as Pumas  have similarly been built using the pure Julia simulation stack as their foundation. Julia's speed is allowing parameter estimation, sensitivity analysis, which require even larger numbers of calls to the ODE solvers.  
 
%When doing parameter estimation or MST :WHAST IS THAT ? virtual population analysis, this same ODE needs to be solve thousands of times, and thus this quickly translates to model analyses taking a day to simulate instead of a month.
 
%MST: I DON'T LIKE THIS AS IT SOUNDS LIKE ADVERTISING: In terms of impact and adoption, it has been noted that these Julia dynamical modeling tools have "emerged as our `go-to' tool for most of our analyses in recent months" from the Director Head of Clinical Pharmacology and Pharmacometrics at Moderna Therapeutics in 2020 \footnote{https://pumas.ai/}. An astute reader will quickly extrapolate the real-world impact that accelerating computational performance of such clinical trial analyses has had.

\section*{Abstraction}

Compared to other programming languages such as R and Python, or C/C++ and Fortran, Julia allows an exceptionally high level of abstraction\cite{Lange2020}. In order to highlight the advantages of an abstract programming language\cite{heroux2020}, we can use an analogy to a more standard tool of a biologist -- a pipettor -- to develop some intuition for the important role of abstraction. 
\par 
The pipettor is a standard piece of lab equipment made with slightly different designs by different manufacturers; nevertheless they all perform the same task in a similar way. It thus takes minimum effort to get used to a new pipettor, without having to retrain on every aspect of an experimental protocol. Abstraction achieves the same for a software. Similar to the described abstract interface ``pipettor'', in Julia we have interfaces such as the {\tt AbstractArray} interface. All implementations of it are “array-like” structures and provide the same core functionalities which an “array-like” structure is expected to have. This allows us to easily swap between different implementations of the same interface and hence promotes specification without the often feared loss of the ability to integrate with existing software or adapt to new applications\cite{Oliveira2006}.
\begin{figure}[h]
    \centering
    \includegraphics[width=0.35\textwidth]{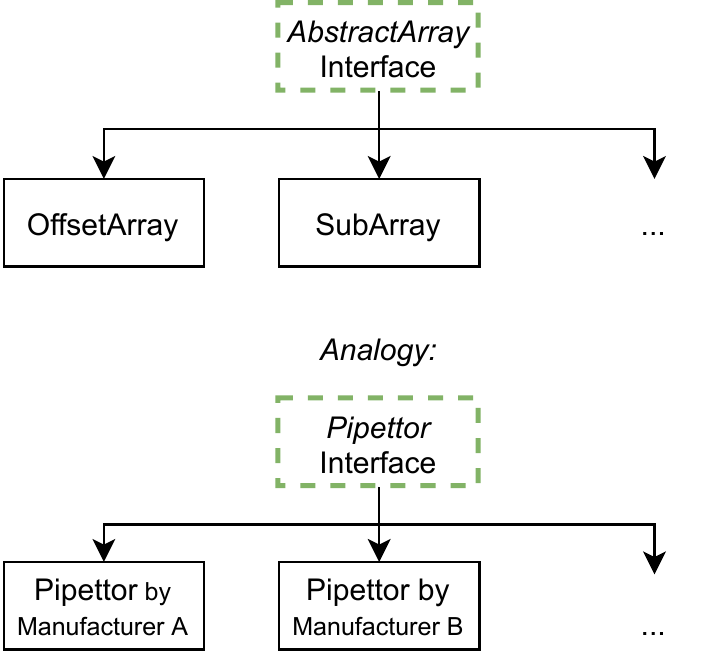}
    \caption{Interfaces: Pipettors and the AbstractArray.}
    \label{fig:abstraction_interfaces}
\end{figure}
\par 
Compared to, e.g., physics, the biological sciences have a high level of heterogeneity in the prevalent data \cite{Alyass2015, GomezCabrero2014, Nagaraj2018}. 
This poses challenges for programming languages\cite{Oliveira2006} and data analysis pipelines, and changes to data may require substantial rewriting of code for processing and analysis. We may end up with different implementations of algorithms for different types of data; or we may remove details and nuance from the data to enable analysis by existing algorithms.
%In a scenario as such, the scientist is given the choice to either use the appropriately complex description and have several versions of the same algorithm for different experiments, or, compromise and remove some of the complexity of the data so that it fits into the same algorithm but potentially loose meaningful details. 
With abstraction we do not have to make such choices. Julia's abstraction capabilities provide room for both specialisation and generalisation through  features such as abstract interfaces and generic functions that can exploit the advantages of unique data formats with varying internal characteristics without an overall performance penalty.
\par 
%Abstraction can also help in the collaborative development of software, or in the integration of software packages developed independently of one another. Julia being open-source and all development hosted on GitHub, provides the backdrop facilating this interplay.
%
%Another side effect of Julia’s abstraction feature is that it enables and encourages collaborative software development. Due to the language design, integrating existing software is easy. And given the fact that Julia is open source and hosted on GitHub, scientist have the whole package ecosystem (Figure \ref{fig:packages} and Table \ref{tab:packages}) and its functionalities at their fingertips. First, this is great for individual scientists. They have access to a whole toolbox of high performance, easy to use, and cutting edge methodology. But from a bigger picture perspective this also means that experts across many fields have easy access to the packages of the ecosystem. This leads to the fact that Julia packages are constantly being tested and peer reviewed. And, as contributions are easy, too (from a technical point of view), the line between user/reviewer and developer blurs quickly and we believe this is the prefect environment for robust and highly optimised methodology and software development.

\begin{table*}[t]
  \centering
  \begin{tabular}{|p{2cm}|p{3.6cm}|p{10.5cm}|}
      \hline
      Community&Topic&Example packages \\
      \hline \hline
      JuliaData&Data manipulation, storage, and I/O& DataFrames.jl, JuliaDB.jl, DataFramesMeta.jl, CSV.jl\\
      \hline
      JuliaPlots&Data visualization&Plots.jl, Makie.jl, StatsPlots.jl, PlotlyJS.jl\\
      \hline
      JuliaStats&Statistics and Machine Learning& Distributions.jl, GLM.jl, StatsBase.jl, Distances.jl, MixedModels.jl, TimeSeries.jl, Clustering.jl, MultivariateStats.jl, HypothesisTests.jl.\\
      \hline
      BioJulia&Bioinformatics and Computational Biology&BioSequences.jl, BioStructures.jl, BioAlignments.jl, FASTX.jl, Microbiome.jl\\
      \hline
      JuliaImages&Image processing&Images.jl, ImageSegmentation.jl, ImageTransformations.jl,
      ImageView.jl\\
      \hline
      EcoJulia&Ecological research&SpatialEcology.jl, EcologicalNetworks.jl, Phylo.jl, Diversity.jl\\
      \hline
      SciML&Scientific machine learning&DifferentialEquations.jl, ModelingToolkit.jl, DiffEqFlux.jl, Catalyst.jl\\
      \hline
      FluxML&Machine Learning&Flux.jl, Zygote.jl, MacroTools.jl, GeometricFlux.jl, Metalhead.jl\\
      \hline
  \end{tabular}
  \caption{Julia provides a rich package ecosystem for biologists. Related packages are organised in package communities. In this table, we present an overview of package communities  we consider most relevant to biologists.}
  \label{tab:packages}
\end{table*}
\begin{figure*}[t]
    \centering
    \includegraphics[width=0.9\textwidth]{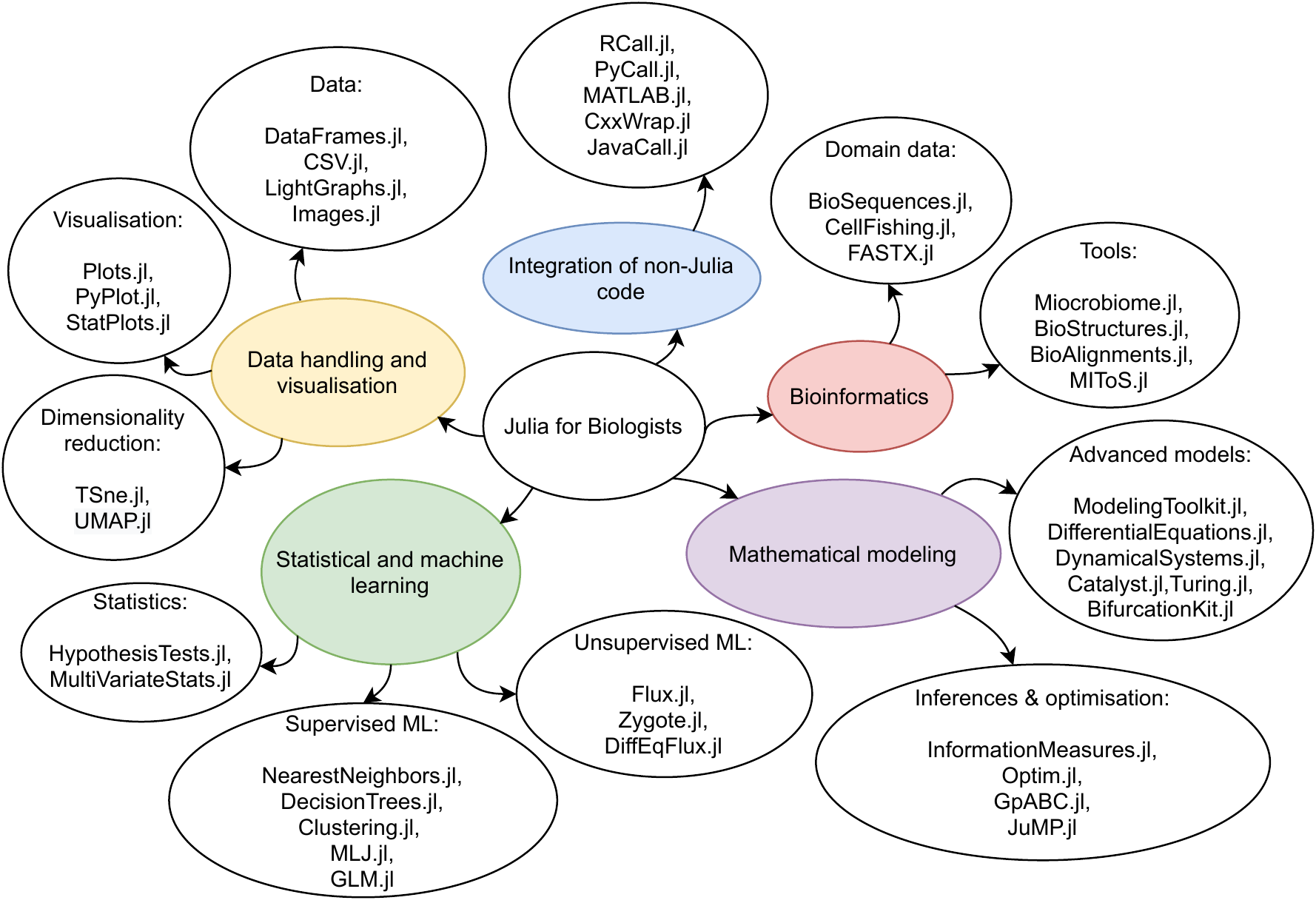}
    \caption{An overview of Julia's package ecosystem presented by topic groups.}
    \label{fig:packages}
\end{figure*}

\paragraph{Example 1: Structural bioinformatics with composable packages.}
The flexibility of Julia means that packages from different  authors can generally be combined easily into workflows, a feature known as composability. Users can benefit from Julia's flexibility just as much as package developers.
For example, we consider a standard structural bioinformatics workflow, where we want to download and read the structure of the protein {\em crambin} from the Protein Data Bank (PDB).
This can be done using the {\tt BioStructures.jl} package \cite{Greener2020} from the BioJulia organisation, which provides the essential bioinformatics infrastructure. Protein structures can be viewed using {\tt Bio3DView.jl}, which uses the {\tt 3Dmol.js} JavaScript library \cite{Rego2014} as Julia can easily connect to packages from other languages.
We can show the distance map of the C\textbeta\ atoms using {\tt Plots.jl}; while {\tt Plots.jl} is not aware of this custom type, a {\tt Plots.jl} recipe makes this straightforward.
{\tt BioSequences.jl} provides custom data types of sequences and allow us to represent the protein sequence efficiently. With this {\tt BioAlignments.jl} can be used to align our sequences of interest. This suite of packages can be used to carry out single cell full-length total RNA sequencing analysis \cite{Hayashi2018} quickly and with ease.
A few lines of code in {\tt BioStructures.jl} allow us to define the residue contact graph using {\tt LightGraphs.jl}, giving access to all the optimised graph operations implemented 
in {\tt LightGraphs.jl} for further analysis, such as calculating the betweenness centrality of the nodes.
If the code is written in {\tt Pluto.jl}, then updating one section updates the whole workflow, which is essential for exploratory analysis.
\par
Packages can be combined to meet the specific needs of each study; for example to generate protein ensembles and predict allosteric sites \cite{Greener2017}, or to carry out information theoretical comparisons using the {\tt MIToS.jl} package \cite{Zea2016}.% and different modules of the MIToS.jl package can be used together to carry out mutual information analysis 
\par
In this example we have used at least five different packages together seamlessly.
{\tt Plots.jl}, {\tt BioAlignments.jl} and {\tt LightGraphs.jl} do not depend on, or know about {\tt BioStructures.jl} but can still be used productively alongside it.
Abstraction means that the improvements in any of these packages will benefit users of {\tt BioStructures.jl}, despite these packages not being developed with protein structures in mind.
\par
This level of package composability is common across the Julia ecosystem and is ultimately enabled by abstract interfaces supported by multiple dispatch, i.e.\ the ability to define multiple versions of the same function with different argument types.
Programmers can define standard functions such as addition and multiplication for their own types, which means that functions in unrelated packages often "just work" despite knowing nothing about the custom types.
This is rarely seen in languages such as Python, R and C/C++, where the behaviour of an object is confined to one place \cite{heroux2020} and combining classes and functions from different projects requires much more (of what is known as) ``boilerplate" code.
For example, the popular Biopython project \cite{Cock2009} has grown over many years to become a powerful package covering much of bioinformatics. But extensions to Biopython objects are generally added to (an increasingly monolithic) Biopython, rather than existing in independent packages.
This can lead to objects and algorithms that have the difficult task of fitting all use cases simultaneously \cite{Kunzmann2018}; this can introduce reservations about adding code that interacts with other packages and/or increases dependencies.
\par
By comparison, the composability of Julia, which is connected to the underlying technical feature of multiple dispatch, enables scientists to combine packages to carry out new tasks in ways probably never imagined by the creators of Julia  \cite{Karpinski2019}.
%This synergy will increase as the Julia package ecosystem grows.
Composability also facilitates writing generic code that can be used beyond its intended application domain.
For example, {\tt Tables.jl} provides a common interface for tabular data, allowing generic code for common tasks on tables; currently, some 129 distinct packages draw on this common core for purposes far beyond the initially conceived application scope.
\begin{figure*}[pt]
    \centering
    \includegraphics[width=\textwidth]{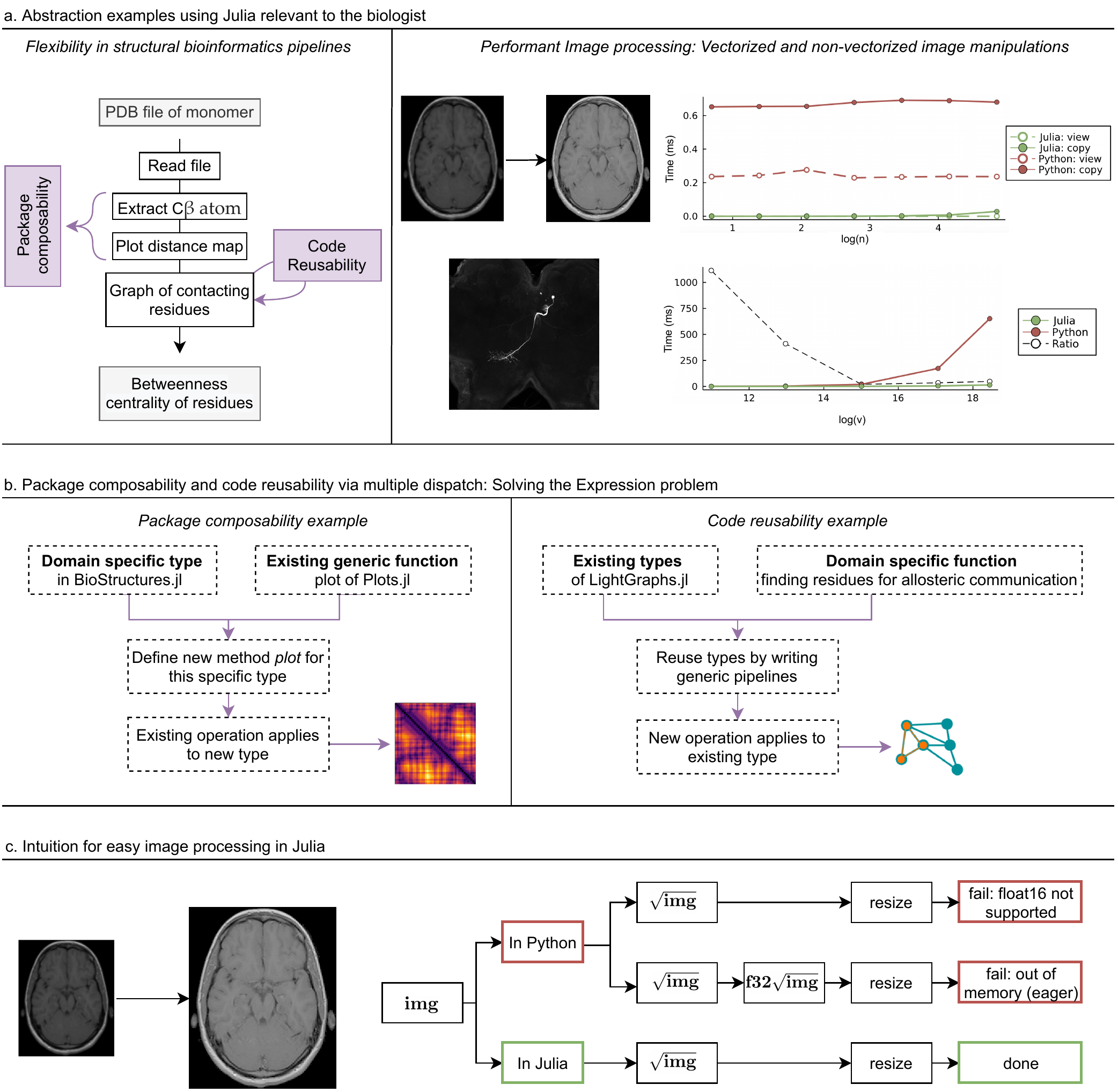}
    \caption{Julia's abstraction feature.  (a) Two examples relevant to the biologist. Left: Package composability and code reusability in structual bioinformatics pipelines. Right: Contrasting (top) and segmenting (bottom) images as examples for high performance vectorizable and non-vectorizable image manipulations in Julia. (b) Conceptual background of package composability and code reusability in structual bioinformatics pipeline: Julia solving the expression problem as it enables an easy addition of types and functionalities without causing compatibility issues with existing code. (c) Intuition for robustness of image processing in Julia vs. Python.}
    \label{fig:abstraction}
\end{figure*}

\paragraph{Example 2: Flexibility and performance in image processing.}
Microscopy in its many forms underlies much of modern biology. But extracting information from imaging data is challenging for two main reasons.
The first challenge lies in the nature of the raw data. Scientific images can be very large, and it is not uncommon for datasets to reach a size of multiple terabytes \cite{Holekamp2008,Tomer2012}. In such instances, initially minor performance inconveniences can quickly extrapolate to become limiting factors for scientific discovery. Also, the images -- typically internally represented as arrays -- often exhibit great diversity: for example, a single imaging dataset may have two or three spatial dimensions, zero or one temporal dimensions, and a color- or modality-channel. This is further exacerbated by the complexity of the accompanying meta information on the imaging conditions and technologies which also influence down stream analysis and interpretation. Having the flexibility to accommodate for this level of diversity whilst also providing the necessary performance needed when dealing with data of this size, is a non-trivial challenge for any programming language with significant implications for outcomes \cite{Schindelin2012,Hofmanninger2020}.
\par 
The second reason for carefully thinking about the choice of software for image processing pipelines, lies in the nature of processing pipelines themselves. Typically the data are sequentially manipulated over multiple steps. In a naive approach, a new, slightly altered version of the large raw dataset is created and stored for each step in the pipeline. This is inherently inefficient and becomes quickly infeasible or impossible as data sizes grow beyond  storage capacities.  Documenting and tracking different combinations of data manipulation steps is also non-trivial as each step could lose information.
%and carries a certain risks of lost precision by itself and therefore might -- even if only unconsciously -- limit the flexibility of the scientist.
\par 
Efficient data representation combined with flexible processing is of essence  to extract meaningful conclusions from the data. Abstraction is Julia's key feature that enables state-of-the-art image processing \cite{Lee2019,Dragomir2019}: by keeping a high level of abstraction in the internal data representations, the diversity in image data can be captured and exploited, and modifications to the data become easier, too. A core component for implementing the relevant abstraction in Julia is provided by the {\tt AbstractArray} interface \cite{AbstractArrays}, especially in its combination with {\em lazy operations}. In a normal, ``eager'', operation, each computation is executed immediately upon being invoked. By contrast, \emph{lazy operations} delay their computations until the latest possible moment in time, i.e. the execution is separated from the declaration of a computation. In some cases, this can be delayed all the way to the moment where we wish to visualize the processed data, so that no computation needs to occur on any data not being inspected.
\par 
As previously described, the {\tt AbstractArray} interface can be thought of as a template which creates an agreement between existing software and the user’s software. We can use the template, i.e. implement the {\tt AbstractArray} interface, in order to define a new array type which optimally fits their data format. By using the interface, we also agree to provide certain functionalities for this new object.
Providing both high levels of customization and standardization of operations that need to be supported by an array-like object, allows for the composition of complex and highly specialised pipelines. The specifics of the array no longer matter because of the abstraction.
Many {\tt AbstractArray} interface implementations helpful in image processing already exist and we do not have to start from scratch for each new imaging modality. Examples include {\tt SubArrays} (region-of-interest ``view" selection), {\tt MappedArrays} (lazy-modification of values), ``ReshapedArrays'' (lazy-modification of dimensionality), and {\tt WarpedViews} in the {\tt ImageTransformations.jl} package (lazy coordinate transformations).
\par
 With effective lazy operations, it becomes possible to manipulate and inspect massive data sets even with relatively modest computing hardware, because the hardware only needs to load, process, and display the small subset of the data being actively explored. Pre-processing stages that might require tuning several parameters to the particulars of the data set can be refined quickly, with each iteration perhaps comprising only a few seconds or minutes, rather than the hours, days, or weeks that might be required if each step had to cached to disk between manipulations.
\par 
Other languages support the concepts of abstraction and lazy operations, too, but despite considerable investment they do not provide the same level of comfort and capability available in Julia. For example, in Python, the most widely-used lazy-operation package is Dask \cite{Daniel:2019ti}, which has a sophisticated engine for managing computational graphs and applying them across distributed data sets. However, when using Dask to process large image data sets, one frequently encounters severe limitations on composability (Figure \ref{fig:abstraction}(c)): some algorithms may not support outputs of previous stages, while others may force an eager intermediate step in the pipeline potentially exhausting memory  resources, and yet others may attempt to allocate an unachievable output array. By contrast, with Julia, one can routinely expect that arbitrary combinations of processing "just work" together, and we can use lazy operations along the whole image processing pipeline. Because Julia aggressively optimizes computations at a granular level (all the way to the single pixel), this flexibility comes with little or no overhead, in marked contrast to languages such as Python (Figure \ref{fig:abstraction}(a) and additional information in the accompanying Git repository).
\par 
%The objective of the software as a tool in the instance of image processing should be to provide the adequate level of flexibility a biologist needs in order to discover new science without being limited by data storage and performance issues caused by the software. Julia and its high level of abstraction provides this flexibility to the biologist and therefore enables new analyses and experiments. For example, Julia has previously been adopted by labs processing large images acquired by light sheet microscopy in mice\cite{Lee2019} and fish\cite{??}. Julia also enabled a real-time two-photon pipeline to perform calcium imaging in intact neural tissue and then select and phototag specific cells that exhibited specific response properties\cite{Lee2019}.
\section*{Metaprogramming} 
\label{sec:metaprogramming}
\begin{figure*}[t]
    \centering
    \includegraphics[width=0.9\textwidth]{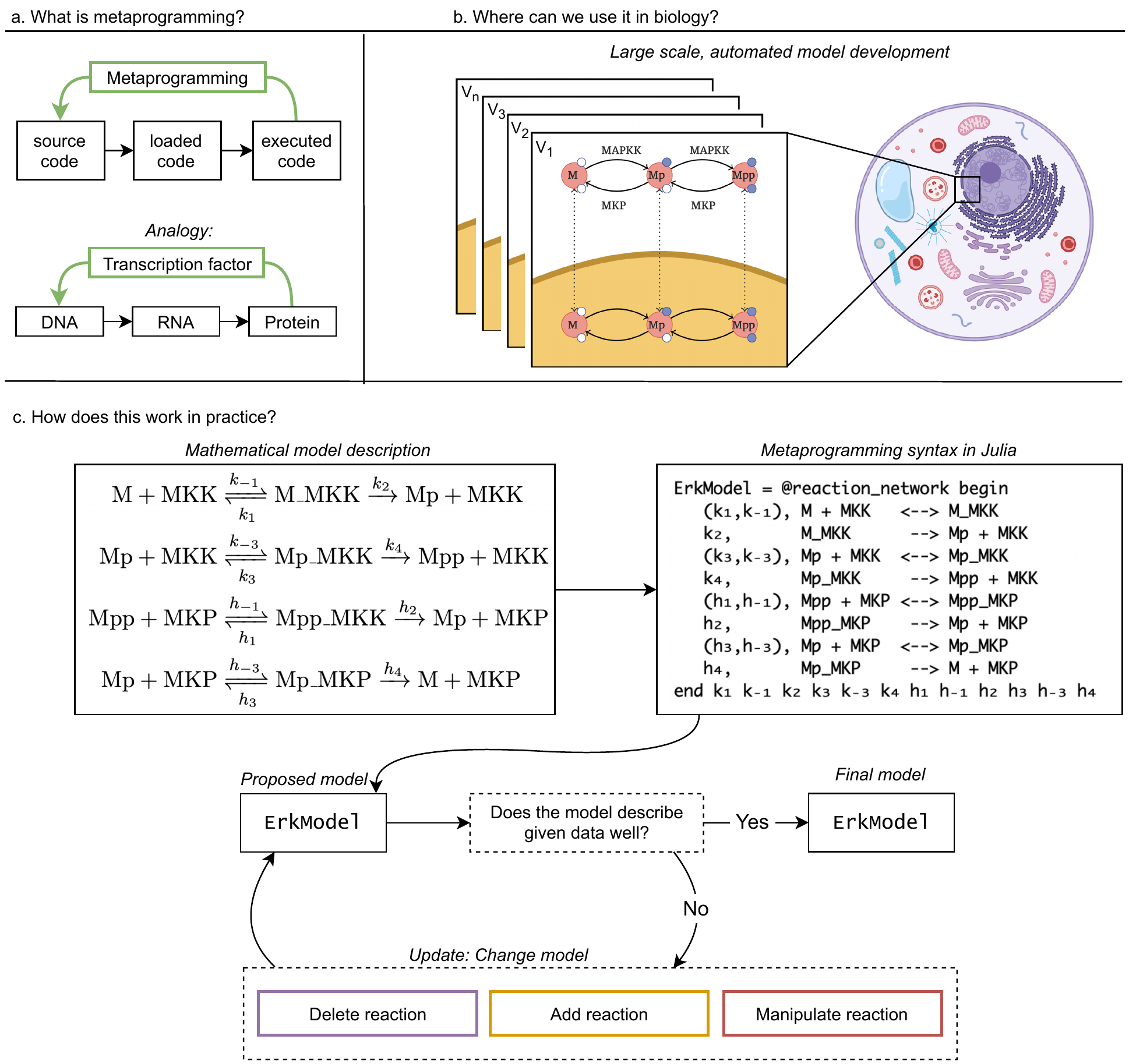}
    \caption{Julia's Metaprogramming feature. (a) Illustration of metaprogramming and an analogy the the central dogma of molecular biology. (b) Application area of metagprogramming in biology. (c) Example workflow.}
    \label{fig:metaprogramming}
\end{figure*}

As our knowledge of the complexities of biological systems grows ever deeper, so does our need of means to simplify the construction and analysis of mathematical models of these systems. Currently, most modeling studies in biology rely on programming languages that treat source code as static: once written, it can be processed into loaded and executing code, but it is never changed during execution of the program. We can compare this linear control process to the central dogma of biology \cite{Crick1970, Hickinbotham11}: Source code (DNA) is transformed into loaded code (RNA), and executing code (protein). In fact, we know now that this process (DNA$\longrightarrow$RNA$\longrightarrow$Protein) is not linear at all: for example, RNA and proteins can alter how and when DNA is translated. Programming languages that support metaprogramming break the linear flow of the computer program in a similar manner to the analogy of the central dogma. With metaprogramming, source code can be written that is processed into loaded and executing code \textit{and} it can also affect the source code. This shifts our perception of code to that of a dynamic instance. By treating code as part of the data, we can write computer code that changes code: the program can modify aspects of itself during run-time (Figure \ref{fig:metaprogramming}(a)).
\par
Metaprogramming enables a form of reflection and learning by the software and the concept originated in early artificial intelligence research, in particular in the context of the LISP programming language. Of course, the ability of a program to modify computer code needs to be channelled very carefully. In Julia, this is done via a feature called {\em hygenic macros} \cite{Perera2008}. These are flexible code templates, specified in the program, and which can be  manipulated at execution time. They are called "hygenic" because they prohibit accidentally using variable names (and thus memory locations) that are defined and used elsewhere.  
These macros can be used to generate repetitive code efficiently and effectively. 
\par
But there are other uses that can enable new research. Perhaps this is most immediately  relevant for modelling biological systems. Unlike in physics, first principles (conservation of energy, momentum, etc., and their related symmetry relationships \cite{Thorne:2017vm}) offer little guidance as to how we should construct models of biological processes and systems. For these notoriously complicated biological systems, trial and error, coupled to biological domain expertise, and state-of-the-art statistical model selection are required at a very minimum \cite{Kirk:2015gj}. Great manual effort is spent on the formulation of mathematical models, the exploration of their behavior, and their adaptation in light of comparisons to data and/or design principles. Metaprogramming, or the abilities of introspection and reflection during runtime \cite{Perera2008},  and the ability to automate these parts of the modeling process open up wide scope for new approaches to modeling biological systems  (Figure \ref{fig:metaprogramming}(b)). 
\par
     
% TO DO: rewrite part "concrete how it’s done in Julia part"

% Example 1: Biochemical reaction networks in Julia: many models from one
\paragraph{Example 1: Biochemical reaction networks.}
Mathematical models of biochemical reaction networks allow us to analyze biological processes and make sense of the bewilderingly complex systems underlying cellular function \cite{Shinar2010, Kirk2013, Warne2019}. But the specification of mathematical models is challenging and requires us to specify all of our assumptions explicitly. 
%additional assumptions are required: are the dynamics deterministic or stochastic? are there time delays? does the law of mass action apply? Based on these assumptions, the resulting model may take on a variety of forms: ordinary differential equations (ODEs), delay differential equations, stochastic differential equations (SDEs), or discrete-time stochastic processes (Gillespie-type models), to give a few examples. To create instances of each of these models would typically -- in an alternative language such as Python or MATLAB -- require writing different blocks of code one for each model to be defined. In Julia, via metaprogamming,  many different models can be generated  automatically from a single block of code.
We then have to solve these models based on assumptions about the dynamics. Solving a given reaction networks can involve solution, for example, of ordinary differential equations (ODEs), delay differential equations, stochastic differential equations (SDEs), or discrete-time stochastic processes. To create instances of each of these models would typically -- in languages such as C/C++ or Python -- require writing different snippets of code for each modelling framework. In Julia, via metaprogamming,  many different models can be generated  automatically from a single block of code.
This simplify workflows and make them more efficient, but also removes the possibility of errors due to model inconsistencies. Good programming tends to be lazy, and the lazyness enabled by metaprogramming reduces the risk of introducing errors into code.
\par 
For example, the ERK phosphorylation process shown in Figure \ref{fig:metaprogramming}(b)\cite{Filippi:2016gs}. Here ERK is doubly phosoporylated (by its cogniscant kinase, MEK) and dephosphorylated by a phosphatase upon which it can shuttle into the nucleus and initiate changes in gene expression. It is a small and important component of a larger signalling network. Its role has made ERK a target of extensive and intensive analysis, and modelling has helped to shed light on its function and role in cell-fate decision  making systems \cite{Michailovici:2014hv}. 
This small system -- albeit one of great importance and subtlety -- forms a building blocks for larger, more realistic biochemical reaction \cite{Shinar2010} and signal transduction \cite{MacLean2015} models.
%
%As an example, consider a simple enzymatic process: an enzyme $E$ binds reversibly with a substrate $S$ to form a complex $C$. While bound, the enzyme converts the substrate into a product $P$, and the enzyme is recovered. Reactions of this form provide the building blocks for large, realistic biological models including those of biochemical reaction networks \cite{Shinar2010} as well as other cell processes such as signal transduction \cite{MacLean2015}. The reactions that comprise this networks are represented as:
%\[E + S \xrightleftharpoons[k_2]{k_1} C \xrightarrow[]{k_3} E + P \] 
%where we assume the law of mass action applies, i.e. the reaction proceeds at a rate proportional to the product of its reactants. 
\par
In Julia, using {\tt Catalyst.jl} \cite{ma2021modelingtoolkit}, this model can be written directly in terms of its reactions, with the corresponding rates \{$k_1,k_2,k_3$\}: source code is human readable and differs minimally from the conventional chemical reaction systems shown in Figure \ref{fig:metaprogramming}(c). 
\par
The science is encapsulated in this little snippet and solving of the reaction systems then proceeds by calling the appropriate simulation tool from {\tt DifferentialEquations.jl}: for a deterministic model specified the reaction network is directly converted into a system of ODEs (via \texttt{ODESystem}). Likewise, the same reaction network can be directly converted into a model that is specified by SDEs (via \texttt{SDEProblem}) or a discrete-time stochastic process model (via \texttt{DiscreteProblem}). Each of these cases
% (achieved through metaprogramming assisted by multiple dispatch)
leads to the creation of a distinct model that can be simulated or analyzed; yet all of the models share the underlying structure of the same reaction network. To simulate one of the resulting models, the user needs to specify only the necessary assumptions required for a simulation -- i.e. the parameter values and initial conditions -- as well as any further assumptions required that are specific to the model type, e.g. the choice of noise model for a system of SDEs. Adapting the model to include nuclear shuttling \cite{Harrington:2013wv} of Erk as in Figure \ref{fig:metaprogramming}(c), or extrinsic noise upstream of Erk \cite{Filippi:2016gs} is easily achieved in the metaprogramming approach.
\par
Fitting models to data, or estimating their parameters from data, is also supported by the Julia package ecosystem. Parameter estimation by optimsing the likelihood, posterior or a cost-function is straightforward using the {\tt Optim.jl} \cite{Mogensen2018} or {\tt JuMP.jl} \cite{Dunning2017} packages. And because of Julia's speed it has become much easier to deploy Bayesian inference methods; here, too, metaprogramming helps tools such as the probabilistic programming environment, {\tt Turing.jl} \cite{Ge2018}. Approximate Bayesian computation approaches are already implemented in Julia in the package {\tt GpABC.jl} \cite{Tankhilevich2020}; they also benefit from Julia's combination of speed, abstraction and metaprogramming, and are faster than implementations in other languages (including some of the authors' work \cite{Liepe:2014iw}).

\paragraph{Example 2: Whole cell modeling.}
An additional application area for metaprogramming is the development of physiologically more realistic models, whether at the levels of whole cells, tissues, or even the physiology of whole organisms. In whole cell modeling, models potentially scale up to the size of $10^3-10^5$ species \cite{Stumpf2021} and a key problem is that constructing models of this size is extremely difficult \cite{Babtie:2017ix,CMason:2019cz}. In fact even small parts of such models, see Figure \ref{fig:metaprogramming}, such as signalling cascades have a large number of (generally unknown) parameters. Here model development cannot rely on manual curation or inspired guesswork \cite{Babtie:2017ix}. Instead automated model development will be required \cite{Stumpf2021}. The reasons for this is that the bookkeeping efforts required to keep track of molecular species, their interactions, and the ways in which molecule numbers change as a result of biochemical interactions, are simply not manageable by conventional means. We do not know the model structure and therefore have to experiment with different model-setups. Without metaprogramming we would have to write or adapt the cellular simulation code for each new attempt. Plus, of course, nobody is able to check the validity of such a large model in the way we can check a simple mathematical model of the type that has traditionally dominated theoretical biology.
\par
Developing a whole cell model will almost certainly involve piecing together sub-models, for which we can build on {\tt Catalyst.jl}. Calibrating such (sub-) models against data -- that is to infer parameters from data -- is a demanding task, that has yet to be solved for such large systems (it is {\em a priori} not clear to what extent this can be solved). Approximations to the dynamics and/or the inference process can help; and for many sufficiently small systems (say signalling networks) current tools will allow us to determine their parameters from literature and/or data, as described in the example above. We may, in addition, want to use efficient approximations to the stochastic dynamics \cite{Lakatos2015,Schnoerr:2017iw}, such as provided by {\tt MomentClosures.jl} \cite{Sukys2021}. This can be coupled to parameter inference, as described above, via optimization ({\tt Jump.jl} \cite{Dunning2017}) or Bayesian inference ({\tt Turing.jl} \cite{Ge2018} and {\tt GpABC.jl} \cite{Tankhilevich2020}).  
\par
 {\tt Catlab.jl} is a package that  makes composing and combining smaller models into a larger model possible, and relatively straightforward. The toolset that we can use to construct such models continues to grow. For example, hypergraphs provide a much more flexible representation for mathematical models that conventional networks, and  For such comprehensive models, grown in a principled way, the {\tt SciML} suite, via e.g. {\tt DiffEqSensitivity.jl} \cite{rackauckas2020universal}, allows us to quantity uncertainty and assess sensitivity of model outputs both locally and globally.  Metaprogramming alleviates the need to "hard-code" such large models. Instead they can be generated automatically without sacrificing the runtime efficiency of the simulation models.
\par
The model development process enabled by the pipeline, Sub-model formulation $\longrightarrow$ Sub-model fitting $\longrightarrow$ Composition of Large Model from Sub-models $\longrightarrow$ Sensitivity and Uncertainty Analysis, 
differs from the way in which the first (and so far only) comprehensive whole cell model was generated, which relied on a lot of manual and expert intervention and input \cite{Karr2012}, which will not scale to other organisms \cite{Stumpf2021}. 
\par
%What is needed are tools that can harvest relevant information from bioinformatics resources and the literature; these can provide baseline scaffolds for whole cell models that can then be further manipulated. For this, for example, tools such as stochastic context free grammars can be employed to propose changes to a model, which are then statistically evaluated in light of available data and compared against alternative models (as exhaustive exploration is impossible for models of this size). To represent models the infrastructure in Catalyst.jl can be used, or we can use hypergraphs (implemented, for example, by SimpleHypergraphs.jl), which offer the flexibility to describe different biomolecular reaction systems.
\par
Overall, metaprogramming in Julia enables the automated construction of models of all sizes: from small biochemical reaction network models to whole cell models. Simulation, inference, and analysis of these models can all be performed with great paucity of code, reducing opportunities for errors to arise, and greatly enhancing our ability to describe and predict complex biological processes with mathematical models. 

\section*{Outlook}
When choosing a programming language we have many choices, but they all boil down to essentially two: do we want to use the language everybody else is using? Or do we want to use the best language for our problem? Traditional languages have an enviable track record of success in biological research. A frightening proportion of the internet  and the modern information infrastructure probably depends on legacy software that would not pass modern quality control. But it does the job, for the moment. Similarly, scientific progress is possible with legacy software. Python, R are far from legacy and have plenty of life in them. And there are tools which allow us to overcome their intrinsic slowness. 
\par
Here we have tried to explain why we consider Julia an attractive language, in our view even the ``best", for the next chapter in the quantitative and computational life-sciences. It is more modern and does not have the ballast of a long track record going all the way to the pre-big-data days. The deliberate choices made by the developers furthermore make it fast and give developers and users of the language a level of flexibility that is difficult to achieve in other common languages such as R, Python, but also C/C++ and Fortran \cite{Sengupta:2019uj}. 
\par
In this work, we have attempted to describe the three main features of Julia’s language design: speed, abstraction and metaprogramming. We have provided some intuition that fills these concepts with life, and we shown in practical terms how they can be exploited in various biological domains, and, in fact, enable new ways of doing biological research. Even though we have introduced these features separately, they are  deeply intertwined. For example, a lot of the speed-up opportunities of Julia derive from the languages abstraction powers; abstraction in turn makes metaprogramming easier. 
\par
These advantages of a new language need to be balanced against the convenience of programmers who are able to tap into the collective knowledge of vast user communities. All languages have started small and had to develop user bases. The Julia community is growing, including in the biomedical sciences; and, it appears to be acutely aware of the needs of newcomers (and underrepresented minorities in the computational sciences more generally) to Julia, which makes the switch to Julia now much easier \cite{Lauwens:2021uj}.

\section*{Acknowledgements}
Thanks to all attendees of the {\em Birds of a Feather} session "Julia for Biologists" at JuliaCon2021. 
Thanks to David F.~Gleich for letting us run an experiment on his servers. 
E.R. acknowledges financial support through a University of Melbourne PhD scholarship. 
A.L.M. acknowledges support from the National Science Foundation (DMS 2045327). 
T.E.H. acknowledges NIH 1UF1NS108176. 
The information, data, or work presented herein was funded in part by ARPA-E under award numbers DE-AR0001222 and DE-AR0001211, and NSF award number IIP-1938400. The  views  and  opinions of authors expressed herein do not necessarily state or reflect those of the United States Government or any agency thereof. 
M.P.H.S. acknowledges funding from the University of Melbourne DRM initiative, and from the Volkswagen Foundation {\em Life?} program grant (grant number 93063). 
\section*{Author contributions} 
E.R. and M.P.H.S. conceived the concept of the project and were in charge of the overall direction and planning. All authors contributed to writing the manuscript, and have read and approved the final version.
\section*{Competing interests}
The authors declare no competing interest.
\bibliography{bib} 
\end{document}